\documentclass[twocolumn,prl,preprintnumbers,amsmath,amssymb,superscriptaddress]{revtex4}
\usepackage{graphicx}
\usepackage{dcolumn}
\usepackage{bm}
\usepackage{ulem}

\font\scripti=cmmi7
\font\scriptscripti=cmmi5
\def\sib#1{\setbox0 = \hbox{\scripti #1}
  \kern-.02em\copy0\kern-\wd0
  \kern.04em\box0} 
\def\ssib#1{\setbox0 = \hbox{\scriptscripti #1}
  \kern-.02em\copy0\kern-\wd0
  \kern.04em\box0} 
\font\tenib=cmmib10 
\skewchar\tenib='177 \skewchar\tenib='177 \skewchar\tenib='177
\def\pbold#1{\setbox0 = \hbox{$ #1 $}
  \kern-.022em\copy0\kern-\wd0
  \kern.011em\copy0\kern-\wd0
  \kern.011em\copy0\kern-\wd0
  \kern.011em\copy0\kern-\wd0
  \kern.011em\box0} 

\usepackage{graphicx}
\usepackage{dcolumn}
\usepackage{bm}
\usepackage{color}

\def\lesssim{\ \raise.3ex\hbox{$<$}\kern-0.8em\lower.7ex\hbox{$\sim$}\ }
\def\gesim{\ \raise.3ex\hbox{$>$}\kern-0.8em\lower.7ex\hbox{$\sim$}\ }

\newcommand{\beginsupplement}{%
        \setcounter{table}{0}
        \renewcommand{\thetable}{S\arabic{table}}%
        \setcounter{figure}{0}
        \renewcommand{\thefigure}{S\arabic{figure}}%
      \setcounter{equation}{0}
        \renewcommand{\theequation}{S.\arabic{equation}}%

     }

\begin{document}
\title{Cooper Triples in Attractive Three-Component Fermions: \\
Implication for Hadron-Quark Crossover}
\author{Hiroyuki Tajima}
\affiliation{Department of Mathematics and Physics, Kochi University, Kochi 780-8520, Japan}
\author{Shoichiro Tsutsui}
\affiliation{Quantum Hadron Physics Laboratory, RIKEN Nishina Center, Wako, Saitama, 351-0198, Japan}
\author{Takahiro M. Doi}
\affiliation{Research Center for Nuclear Physics (RCNP), Osaka University, 567-0047, Japan}
\affiliation{RIKEN iTHEMS, Wako, Saitama, 351-0198, Japan}
\author{Kei Iida}
\affiliation{Department of Mathematics and Physics, Kochi University, Kochi 780-8520, Japan}
\date{\today}
\begin{abstract}
We investigate many-body properties of equally populated three-component fermions 
with attractive three-body contact interaction in one dimension.
A diagrammatic approach suggests the possible occurrence of Cooper triples at low temperature, 
which are three-body counterparts of Cooper pairs with a two-body attraction.  
We develop a minimal framework that bridges the crossover from tightly-bound trimers to Cooper triples 
with increasing chemical potential
and show how the formation of Cooper triples occurs in the grand-canonical phase diagram.
Moreover, we argue that this non-trivial crossover is similar to the hadron-quark crossover proposed in dense matter.
A coexistence of medium-induced triples and the underlying  Fermi sea at positive chemical potential is analogous to quarkyonic matter consisting of baryonic excitations and the underlying quark Fermi sea.
The comparison with the existing quantum Monte Carlo results implies that the emergence of these kinds of three-body states 
can be a microscopic origin of the peak of the sound velocity along the crossover.
\end{abstract}
\maketitle
\noindent {\it Introduction---}
The Cooper problem, where two-component fermions with a two-body attraction undergo an instability toward superconductivity, 
{brought about} a significant breakthrough in condensed matter and particle physics~\cite{BCS}.
{On the other hand}, three-body and higher-body interactions occurring among particles with internal degrees of freedom {play a significant role in cold atomic and nuclear physics}~\cite{Hammer,Carlson,APR,Baym,Fukushima}.

In ultracold atoms, 
{the importance of the residual three-body interaction in a one-dimensional {(1D)} system~\cite{Mora,Mazets} and resulting trimer formation~\cite{pricoupenko2} have been pointed out.}
{Moreover, }{not only} the realization of {non-negligible multi-body interactions}~\cite{Petrov1,Petrov2,Guijarro,Pricoupenko},
{but also} various related phenomena have been proposed~\cite{Sekino,Valiente2,Valiente,Harshman,Nishida1,Horinouchi}.
{Recently, the conditions for attractive and repulsive three-body interactions~\cite{Pricoupenko3} and the Bose-Fermi duality including three-body forces have been discussed~\cite{Valiente3,Valiente4,Sekino2}.
}
\par
\begin{figure}[t]
\begin{center}
\includegraphics[width=4cm]{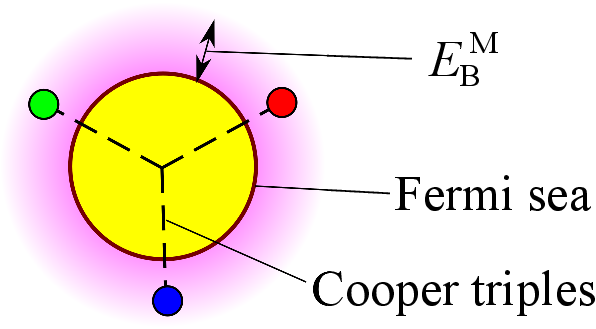}
\end{center}
\caption{
{Schematic figures for 
{the} Cooper triple {phase} in {momentum} space.
{The system exhibits a coexistence of the underlying Fermi sea and loosely bound trimers, that is, Cooper triples, near the Fermi surface with the small width of the energy shell typically given by the in-medium trimer binding energy $E_{\rm B}^{\rm M}$.
Although we {work in one dimension,} we show higher dimensional configurations for visibility. Note that the Cooper triple phase has been predicted in three dimensions~\cite{Niemann,Tajima2021}.
}}}
\label{fig1r}
\end{figure}

{Other interesting aspects of the three-body interaction are the emergences of a quantum scale anomaly and an asymptotic freedom} in 
{non-relativistic {1D} three-component fermions}~\cite{Drut1}.
{In fact, such a} system possesses scale invariance classically~\cite{PitaevskiiRosch,Bergman}, 
while this scale invariance is broken {by the presence} of
three-body quantum bound states.  
{
This anomaly is associated with the asymptotic freedom according to which the running coupling constant becomes progressively weaker in a high-energy regime as in quantum chromodynamics (QCD)~\cite{Schafer}.
}
The same anomaly {also emerges in} two-dimensional {(2D)} fermions with two-body attraction~\cite{Olshanii,Hofmann,Hu,Yin,Vogt,Holten,Peppler,Murthy}. 
At {low density},
the molecular bosonic condensate has been observed in the {2D} system~\cite{Murthy2},
{while a gas of} Fermi degenerate trimers {is expected to be} realized in {the {1D} system}~\cite{Drut1,Daza,Maki,McKenny1,McKenny2}. 
Even at {high density}, the 2D system undergoes
a Cooper-pair {condensation.} 
In the {1D} system, however, 
{the three-body counterparts {of Cooper pairs remain
to be} explored.}  
{A candidate is a Cooper triple (see Fig.~\ref{fig1r}) predicted in 3D three-component Fermi gases with two-body attraction~\cite{Niemann,Tajima2021}.}
It is important to see the stability of such an exotic state, 
{given that} the medium effect on {Efimov} trimer states {is non-negligible}~\cite{MacNeill,Niemann,Nygaard,Sun,Tajima1,Sanayei,Tajima2021}.
\par
{In {dense} QCD, {moreover,} Cooper triples {might be relevant to the hadron-quark continuity~\cite{BaymR,KojoR} because} 
{quarks are three-component fermions in color space.}
{
So far, various scenarios have been discussed in connection with recent astrophysical observations.
One of the intriguing state is}
quarkyonic matter~\cite{McLerran1}, which has been proposed to describe the intermediate-{density} regime {as a state in which} quark and baryonic degrees of freedom coexist {in the course of} the hadron-quark crossover~\cite{BaymR,KojoR} {where} typical energy-scale separations {occur} among the Debye screening mass $m_{\rm D}$, QCD energy scale $\Lambda_{\rm QCD}$, and quark chemical potential $\mu_{q}$ as $m_{\rm D}\ll \Lambda_{\rm QCD}\ll \mu_q$~\cite{McLerran1,Hidaka}.
{Another interesting picture called percolation has also been proposed, where
quark deconfinement starts with formation of a percolation network~\cite{Baym,Fukushima}.} 
While such states have been investigated phenomenologically and the resulting equation of state is consistent with recent astrophysical observations of neutron stars~\cite{McLerranReddy,FukushimaKojo,Kojo,Fukushima},
{a microscopic mechanism of these many-body phenomena is not obvious even at a qualitative level.}}
{Thus, it will be interesting if there is a connection between baryonic excitations in dense QCD and possible color Cooper triples.}
\par
In this work, 
as a quantum simulator of the hadron-quark crossover,
we address many-body properties of 1D three-component fermions with a three-body attraction, {where a quantum Monte Carlo (QMC) simulation has been performed recently~\cite{McKenny2}}.
{{As we shall see,} the Cooper triple phase {occurs at sufficiently large} fermion chemical potential, {i.e.,} $\mu\gesim E_{\rm B}$ {with the
in-vacuum trimer binding energy} $E_{\rm B}$, {to ensure} the coexistence of the Fermi sea and loosely-bound triple states, which
is distinct from a simple large-trimer gas conjectured in Ref.~\cite{Drut1}.
Although mesons have lighter masses ($\simeq 140$~MeV) than baryonic ones ($\simeq 940$~MeV),
dense quark (or quarkyonic) matter is dominated by quark and baryonic degrees of freedom 
due {to the Pauli blocking being effective at sufficiently} large $\mu_q$, 
{leading to a similarity} to the present model with the three-body attraction.}
{Moreover, the existence of three-quark attraction has been revealed by the lattice QCD~\cite{Takahashi1,Takahashi2} and associated Y-shaped color-flux distributions have also been found~\cite{Ichie1,Ichie2}.}

\begin{figure}[t]
\begin{center}
\includegraphics[width=7cm]{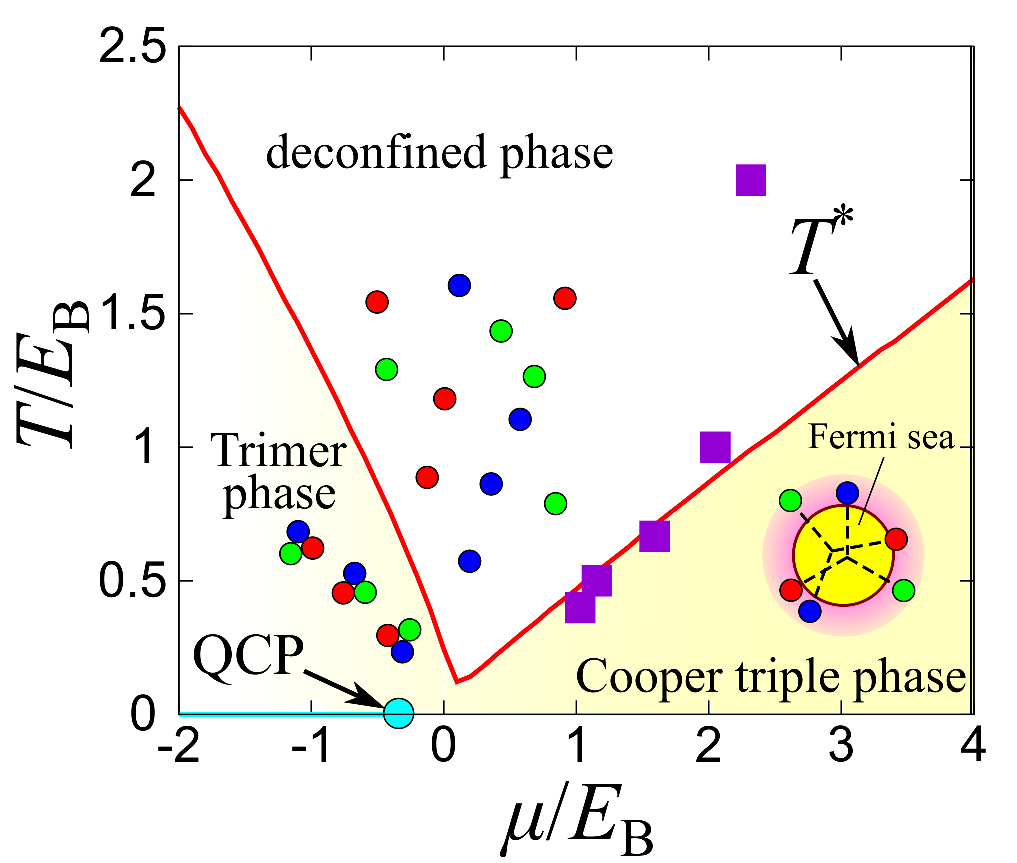}
\end{center}
\caption{
Grand-canonical phase diagram of one-dimensional three-component fermions with a three-body attraction. 
$T^*$ is the temperature where the in-medium trimer binding energy $E_{\rm B}^{\rm M}$ disappears.
$\mu=-E_{\rm B}/3$ at $T=0$ is a trivial quantum critical point (QCP) for the transition {from a zero-density (vacuum) to nonzero-density state.}
{The purple squares show {the points where the isothermal compressibility exhibits a minimum as a function of $\mu$ in the} 
QMC results~\cite{McKenny2}.}
}
\label{fig1}
\end{figure}

\noindent {\it Short summary---} Analyzing three-body spectra, we construct the grand-canonical phase diagram as shown in Fig.~{\ref{fig1}}.
We demonstrate that while the characteristic temperature $T^*$ for {the in-medium three-body state} is 
suppressed 
by thermal {agitation~\cite{Tajima1} around {$\mu=0$},} it linearly increases with the chemical potential in the high-density regime, 
indicating the importance of the Fermi surface effect.
Such different tendencies {between the two regimes lead to} the nontrivial crossover from {the} tightly bound trimer state to Cooper triple phase, which is a three-body counterpart of the BCS to Bose-Einstein condensation (BEC) crossover in two-component Fermi gases~\cite{Zwerger,Randeria,Strinati,Ohashi}.
{The Cooper triple phase is characterized by three-body correlations near the atomic Fermi surface, which is analogous to baryonic excitations in quarkyonic matter.} 
{Although the present system does not involve {gauge fields,} the crossover from bound trimers to Cooper triples {is} reminiscent of the hadron-quark crossover in QCD.}

\par
\noindent {\it Formalism---}
We start from a Hamiltonian $H$ for non-relativistic three-component fermions with a three-body {force} {in {1D}}:
\begin{align}
\label{eq:H}
H&=\sum_{\gamma={\rm r,g,b}}\sum_{{p}}\xi_{{p},\gamma}c_{{p},\gamma}^\dag c_{{p},\gamma}\cr
&+g_3\sum_{{P},{k},{q},{k}',{q}'} c_{\frac{{P}}{3}+{k}-\frac{{q}}{2},{\rm r}}^{\dag}c_{\frac{{P}}{3}+{q},{\rm g}}^\dag c_{\frac{{P}}{3}-{k}-\frac{{q}}{2},{\rm b}}^\dag \cr
&\quad\times c_{\frac{{P}}{3}-{k}'-\frac{{q}'}{2},{\rm b}}c_{\frac{{P}}{3}+{q}',{\rm g}}c_{\frac{{P}}{3}+{k}'-\frac{{q}'}{2},{\rm r}},
\end{align}
where $\gamma={\rm r,g,b}$ {denote the internal degrees} of freedom of fermions,
$\xi_{{p},\gamma}={p}^2/(2m_\gamma)-\mu_\gamma$ is the kinetic energy of a fermion with momentum ${p}$ and mass $m_\gamma$, 
measured {with respect to the} chemical potential $\mu_\gamma$, {and}
$c_{{p},\gamma}^{(\dag)}$ is {the} fermionic annihilation (creation) operator.
The second term in Eq.~(\ref{eq:H}) denotes the three-body interaction with a contact-type coupling constant $g_3$,
{taken to be negative here.}
{We note that the manipulation of $g_3$ has theoretically been proposed in cold atomic~\cite{Petrov1,Petrov2,Guijarro,Pricoupenko,Akagami} and in Rydberg atomic systems~\cite{Gambetta,Ornelas}. 
In {Supplemental} Material S1~\cite{SupMat}, we present one of the possibilities of experimentally realizing this interaction} 
in an atom-trimer resonance model {by analogy with} the optical Feshbach resonance~\cite{Chin} in connection with a closed-channel trimer state~\cite{CvitasM,GhassemiM,YanM,HorikoshiM} and optical control methods~\cite{Wu1M,Wu2M,JagannathanM,Arunkmar1M,Arunkmar2M}.
{For example, applying such a method to an existing mixture (e.g., $^{173}${Yb}) with negligibly small two-body interactions away from the Feshbach resonance enables us to obtain a system with the dominant three-body interaction. }
\par
Many-body effects {are incorporated via} the in-medium three-body $T$-matrix $T_{3}^{\rm MB}({P},i\Omega_n)$, where {${P}$ is the center-of-mass momentum and} $\Omega_n=(2n+1)\pi T$ is {the} fermion Matsubara frequency {with} $n\in\mathbb{Z}$. 
The explicit form of $T_{3}^{\rm MB}({P},i\Omega_n)$ reads
\begin{align}
T_3^{\rm MB}({P},i\Omega_n)=\left[\frac{1}{g_3}-\Xi({P},i\Omega_n)\right]^{-1},
\end{align}
where 
\begin{align}
\label{eq:xi}
\Xi({P},i\Omega_n)
&=\sum_{{k},{q}}\frac{F({k},{q},{P})}{i\Omega_n-\xi_{\frac{{P}}{3}+{k}-\frac{{q}}{2},{\rm r}}-\xi_{\frac{{P}}{3}+{q},{\rm g}}-\xi_{\frac{{P}}{3}-{k}-\frac{{q}}{2},{\rm b}}}.
\end{align}
The statistical factor $F({k},{q},{P})$ in Eq.\ (\ref{eq:xi}) is given by
\begin{align}
\label{eq:F}
F({k},{q},{P})&=\bar{f}_{\frac{{P}}{3}+{k}-\frac{{q}}{2},{\rm r}}\bar{f}_{\frac{{P}}{3}+{q},{\rm g}}\bar{f}_{\frac{{P}}{3}-{k}-\frac{{q}}{2},{\rm b}}\cr
&+{f}_{\frac{{P}}{3}+{k}-\frac{{q}}{2},{\rm r}}{f}_{\frac{{P}}{3}+{q},{\rm g}}{f}_{\frac{{P}}{3}-{k}-\frac{{q}}{2},{\rm b}},
\end{align}
{with} the Fermi-Dirac distribution function $f_{{k},\gamma}=(e^{\xi_{{k}}/T}+1)^{-1}$ and $\bar{f}_{{k},\gamma}=1-f_{{k},\gamma}$.
While {preceding works allow for} the Pauli-blocking effect for the two-body sector~\cite{Niemann} only via the first term in Eq.~(\ref{eq:F}) {that
has} the Fermi momentum $k_{\rm F}$ introduced as a momentum cutoff at $T=0$, the second term, {which represents} three-hole excitations, is also important at finite temperature~\cite{Tajima1,TajimaJLTP}.
By taking $F({k},{q},{P})=1$, one can reproduce {the in-vacuum} three-body $T$-matrix $T_3({P},\Omega_+)$ {that appears} in a three-body problem.
{We note that the resummation of specific ladder diagrams for attractive interactions works well even in 1D at finite temperature~\cite{TTD}.} 
\par
We are interested in {the} conditions that {allow} a trimer {to appear} {in a medium.}
While the three-body binding energy {$E_{\rm B}=\frac{\Lambda^2}{m}e^{\frac{2\sqrt{3}\pi}{mg_3}}$} corresponds to the negative energy pole $\Omega=-E_{\rm B}$ {of the {in-vacuum} three-body $T$-matrix $T_3({P}=0,\Omega_+)$~\cite{note1} {(see also Supplemental Material S2~\cite{SupMat})},} where $\Omega_+=\Omega+i\delta$ involves an {infinitesimally} small imaginary part $i\delta$ {with $\delta>0$}, 
the in-medium {binding} energy $E_{\rm B}^{\rm M}$ can be obtained from 
\begin{align}
\frac{1}{g_3}-\Xi({P}=0,\Omega=-E_{\rm B}^{\rm M}-3\mu)=0.
\end{align}
{Note that a Cooper triple can be defined as a state in which the corresponding pole energy $\Omega+3\mu=-E_{\rm B}^{\rm M}$ is negative at positive $\mu$ and its absolute value is also smaller than $3\mu$.
This is why the regime $E_{\rm B}^{\rm M}\ll \mu$ is consistent with the presence of Cooper triples.}
\par
\noindent {\it Results and discussions---}
{Let us focus hereafter} on symmetric three-component fermions ($m\equiv m_{\rm r}=m_{\rm g}=m_{\rm b}$ and $\mu\equiv\mu_{\rm r}=\mu_{\rm g}=\mu_{\rm b}$) 
in {one} dimension.  
We determine the temperature $T^*$ where $E_{\rm B}^{\rm M}$ disappears; {the result is} shown in Fig.~\ref{fig1}.
Although $T^*$ does not {imply the presence of any kind of phase transition}, it is {still worth knowing.}
Indeed, $T^*$ is qualitatively equivalent to the mean-field critical temperature {that can be regarded in the context of} the BCS-BEC crossover
as the temperature where preformed Cooper pairs appear {incoherently} due to the strong two-body attraction~\cite{Zwerger,Randeria,Strinati,Ohashi}.  
In the numerical calculation we take $E_{\rm B}^{\rm M}=10^{-2}E_{\rm B}$ and $\delta=10^{-3}E_{\rm B}$ since $\Omega=0$ has a singularity due to the edge of continuum.
{We confirmed that our {estimate of $T^*$ is practically} unchanged for smaller $\delta$.} 
\par
\begin{figure}[t]
\begin{center}
\includegraphics[width=8.5cm]{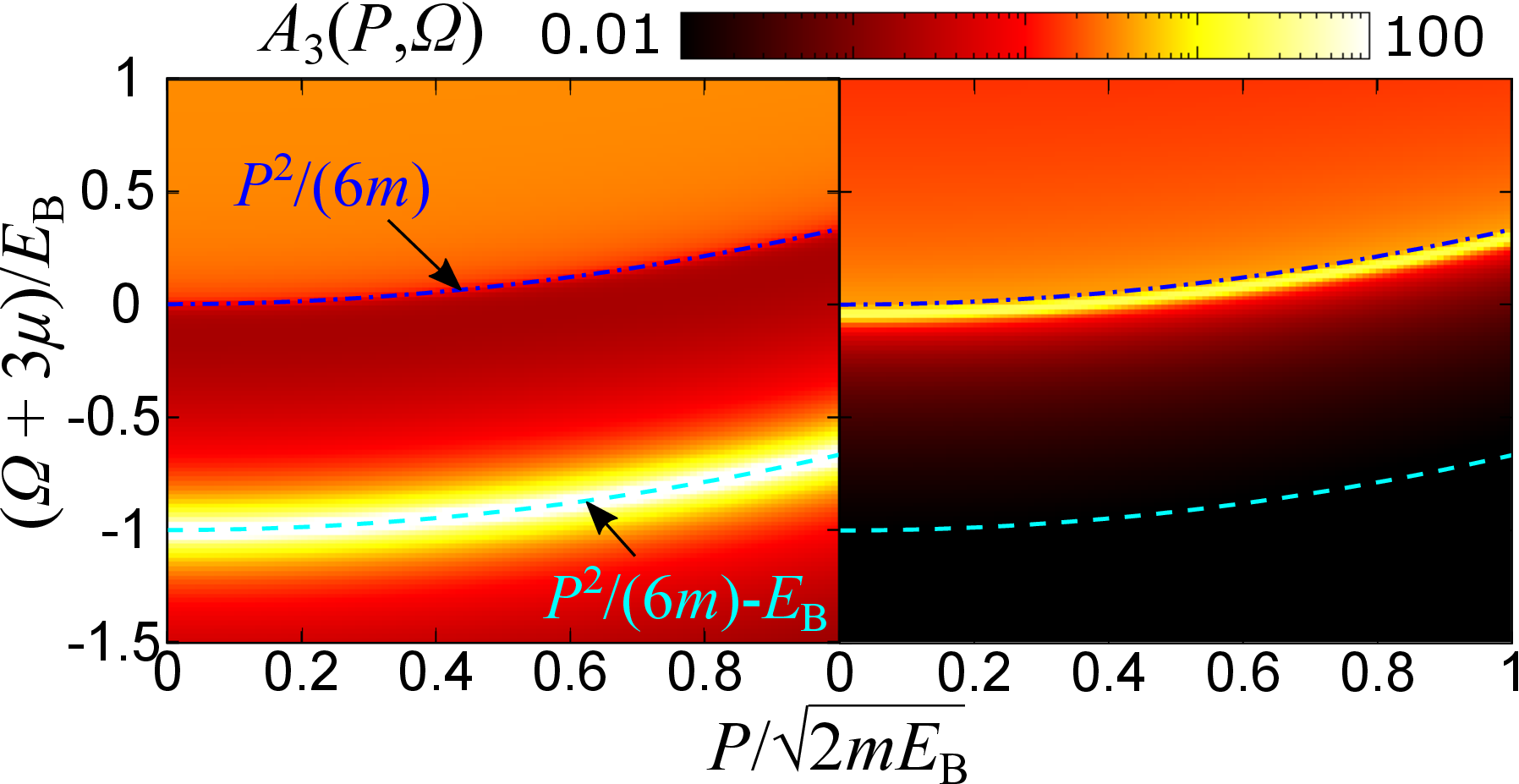}
\end{center}
\caption{Three-body spectral functions $A_3(P,\Omega)$ {calculated} at (a) $\mu/E_{\rm B}=-1$ and (b) $\mu/E_{\rm B}=2$.
The temperature is set at $T/E_{\rm B}=0.1$ in each panel.}
\label{fig3}
\end{figure}
\par
{Let us turn to} in-medium three-body properties at low temperature.
{In Fig.~\ref{fig3},} we {display} the three-body spectral function $A_3(P,\Omega)=-{\rm Im}T_3^{\rm MB}(P,i\Omega_n\rightarrow\Omega_+)$ {calculated} at $T/E_{\rm B}=0.1$. 
{Naturally, the} medium effect is not significant at low density~\cite{Saha,Langmuir}.  
{In fact,} as {depicted} in Fig.~\ref{fig3}(a) {for a typical dilute condition like} $\mu/E_{\rm B}=-1$,
$A_3(P,\Omega)$ has a strong intensity around the dispersion of {a} tightly bound trimer given by $\Omega=P^2/(6m)-E_{\rm B}-3\mu$, {as well as} a continuum above $\Omega=P^2/(6m)-3\mu$.
{The higher the density, the stronger the medium effect.
Consequently,} as shown in Fig.~\ref{fig3}(b), the bound-state peak in $A_3(P,\Omega)$ is strongly suppressed {at $\mu/E_{\rm B}=2$.}  
This {suppressed} {peak, however,} does not merge {into} the continuum at {sufficiently} low temperature.  {Instead,}
the {$P=0$} bound-state pole $\Omega=-E_{\rm B}^{\rm M}-3\mu$ remains {just} below $\Omega=-3\mu$, {which implies} the existence of {an} in-medium trimer {near the Fermi surface ($0<E_{\rm B}^{\rm M}\ll\mu$)}, that is, {a} Cooper triple.  
\par
{We remark that while a} molecular state competes with {a} Cooper triple state in the case of three-component Fermi gases with two-body interactions at finite temperature~\cite{Tajima1}, Cooper triples are not suppressed by such an effect in our model without two-body interactions. 
{Even in the present case, {however,}} 
 Cooper pairs {may} {occur, e.g.,} due to the effective two-body coupling $g_2^{\rm eff}=g_3\rho_{{\rm r}}$ between fermions with $\gamma={\rm g}$ and ${\rm b}$ {($\rho_{\rm \gamma}$ is the number density of $\gamma$ component).}
{This coupling, which} may involve a two-body bound state {of binding energy} $E_{{\rm 2b}}=m(g_2^{\rm eff})^2/4$ in {1D},
is irrelevant {for} a large $\Lambda$ since $g_3$ and {hence} $g_{2}^{\rm eff}$ behave as $\sim 1/\ln(mE_{\rm B}/\Lambda^2)$.
{We remark in passing that in the case of finite-range three-body interactions, irrespective of whether attractive or repulsive~\cite{Pastukhov}, $g_{2}^{\rm eff}$ can be finite and plays a significant role for the interplay between two-body and three-body correlations.}
\par
We {also} note that {while a} trimer-trimer pairing state {used to be invoked} as {one of the possible} ground {states} in Ref.~\cite{Drut1}, the trimer-trimer interaction, {which was later} found to be repulsive~\cite{McKenny1}, would keep Cooper triples unpaired. 
The repulsive trimer-trimer interaction may lead to the trimer Luttinger liquid {(TLL)} in the low-density regime at sufficiently low temperature~\cite{Guan}.
Our results imply the crossover from the gapless excitation in {TLL} to the collective mode of Cooper triples with increasing $\mu$. 
Indeed, a similar crossover of the sound mode has been reported in the {1D} BCS-BEC crossover~\cite{Recati2004}. 
Finally, we emphasize that our prediction of $T^*$ properly allows for thermal agitation, to which the {TLL} picture is in turn susceptible.
From the recent study on finite-temperature Luttinger liquids~\cite{He2020}, one may expect the crossover from {TLL} to the normal trimer or Cooper triple phase with increasing $T$ below $T^*$.
\begin{figure}[t]
\begin{center}
\includegraphics[width=8.5cm]{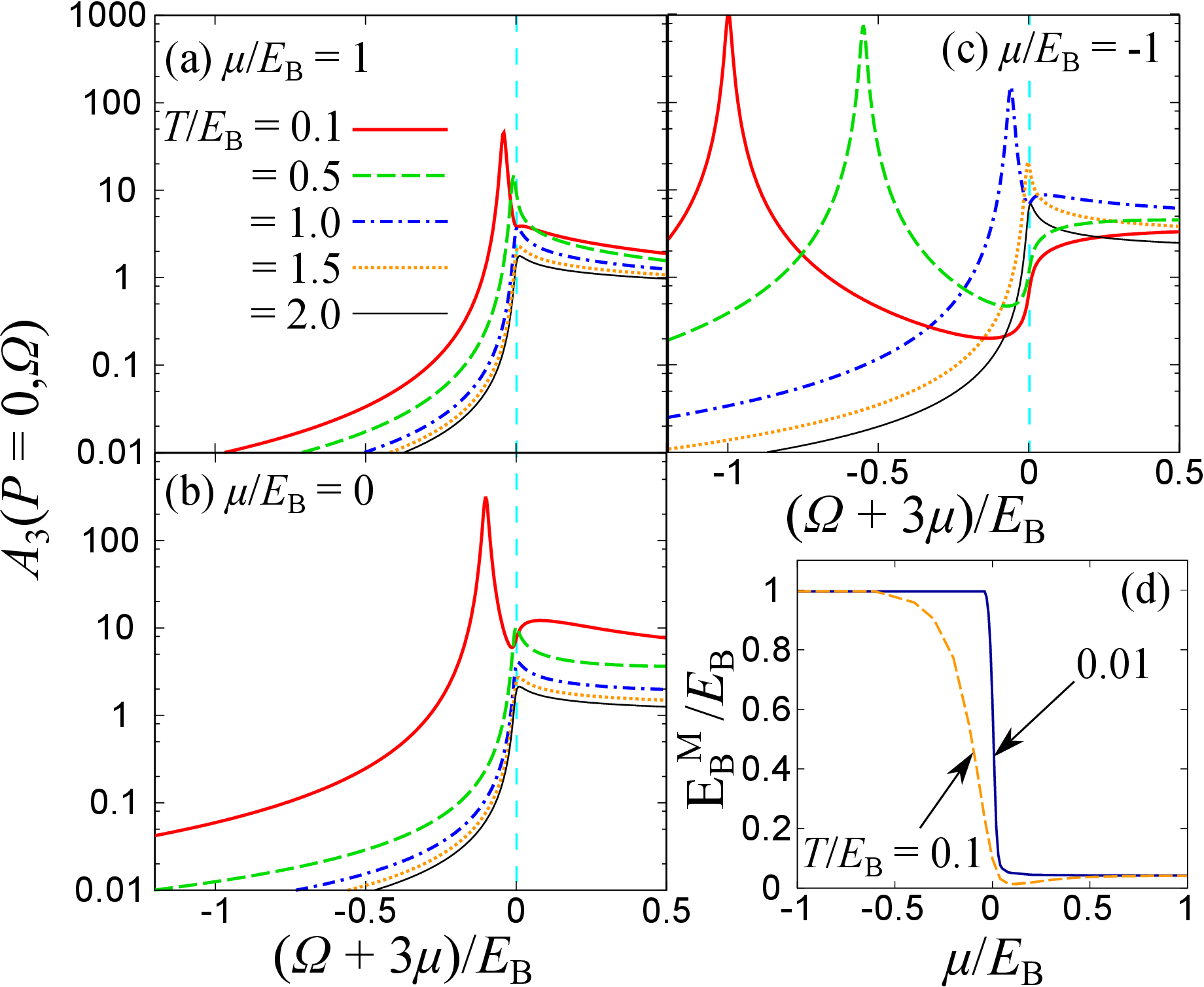}
\end{center}
\caption{Temperature dependence of the three-body spectral functions $A_3(P=0,\omega)$ at (a) $\mu/E_{\rm B}=1$, (b) $\mu/E_{\rm B}=0$, and (c) $\mu/E_{\rm B}=-1$.  
The panel (d) shows the in-medium trimer binding energy $E_{\rm B}^{\rm M}$ as a function of $\mu$ at $T/E_{\rm B}=0.01$ and $T/E_{\rm B}=0.1$.
}
\label{fig4}
\end{figure}
{Figures~\ref{fig4}(a)--(c) show} the three-body spectral functions $A_3(P=0,\Omega)=-{\rm Im}T_3^{\rm  MB}(P=0,i\Omega_n\rightarrow\Omega_+)$ at different temperatures.
Even {in the case of} positive chemical potential $\mu/E_{\rm B}=1$ {depicted in Fig.~\ref{fig4}(a),}  
{a} bound-state peak {occurs just below $\Omega+3\mu=0$} at $T/E_{\rm B}=0.1$.
With increasing temperature, the bound state pole {approaches zero energy} 
and {eventually} merges with the continuum at $T=T^*$, {which amounts to $\sim0.5E_{\rm B}$} at $\mu/E_{\rm B}=1$.
At $\mu/E_{\rm B}=0$, {as shown} in Fig.~\ref{fig4}(b),
the bound-state {peak at $T/E_{\rm B}=0.1$} is {located at a lower energy and 
also enhanced as} compared to the case {of} positive $\mu$ {depicted} in Fig.~\ref{fig4}(a).
{At higher temperature, however,} the pole {at $\mu/E_{\rm B}=0$ is vulnerable to thermal agitation}
and {hence} $T^*$ has a minimum around $\mu=0$.
At {sufficiently} low density, {$\mu/E_{\rm B}$ becomes negative.  For example,
in the case of $\mu/E_{\rm B}=-1$ depicted in Fig.~\ref{fig4}(c),} the bound-state pole 
{reduces} to $-E_{\rm B}$ at {sufficiently} low temperature, {a behavior consistent with} Fig.~\ref{fig3}(a).
This is a clear evidence of {the presence of} tightly bound trimers.
When the temperature increases, the bound-state pole {approaches zero energy} again 
and finally disappears even {in such a low-density regime.}
Since the system does not form the Fermi surface at negative $\mu$, 
{such a} reduction of the trimer binding energy {has to be} associated with thermal agitation~\cite{Tajima1}.
\par
In Fig.~\ref{fig4}(d), we show {how} $E_{\rm B}^{\rm M}$ {evolves} from the tightly bound trimer phase to the Cooper triple phase 
at $T/E_{\rm B}=0.01$ and $0.1$.
{One can see a dramatic drop of $E_{\rm B}^{\rm M}$ around $\mu=0$, indicating the change of {the} system's properties.}
{Both} at {such} low temperatures, $E_{\rm B}^{\rm M}$ continuously changes from {$E_{\rm B}$} to the Cooper triple energy $E_{\rm CT}/E_{\rm B}\simeq 0.04$ with increasing $\mu$. 
Here we note that at exactly zero temperature there is a trivial quantum critical point at $\mu/E_{\rm B}=-1/3$ for the transition from {a zero-density (vacuum) 
to nonzero-density state.}
{Note that this critical point is characterized by the effective fugacity $z_{\rm eff}=e^{(3\mu+E_{\rm B})/T}$ of a bound trimer~\cite{Ngamp}, as $z_{\rm eff}$ becomes exactly zero at $T=0$ when $3\mu< -E_{\rm B}$. In the absence of $E_{\rm B}$, this transition would occur at $\mu=0$.}
From comparison between {the} results of $T/E_{\rm B}=0.01$ and $T/E_{\rm B}=0.1$, one can see that {thermal agitation} becomes significant around $\mu/E_{\rm B}=-1/3\sim 0$.
{All these low-temperature properties}
reflect {the fact} that while the competition between the {three-body} binding and the thermal agitation, {which is} characterized by the ratio $T/E_{\rm B}$, {{manifests} itself} in the low-density regime ($\mu\lesssim 0$),
{the formation of {Cooper triples} in the high-density regime ($\mu \gg E_{\rm B}$) is robust against the thermal agitation due to the Fermi surface effect.}
{The Cooper triple phase {in the high-density regime} can be identified by a typical energy separation $E_{\rm B}^{\rm M}\ll E_{\rm B}\lesssim \mu$, {which is analogous to that in quarkyonic matter.} 
In the low-density regime, see Supplemental Material S3~\cite{SupMat}.} 
\par
{We finally revisit the $\mu$ dependence of $T^*$ shown in Fig.~\ref{fig1}.}  
In {the high-density} regime, $T^*$ linearly increases with increasing $\mu$.
Indeed, this behavior is well fitted by the linear function $T^*=0.384\mu+0.095E_{\rm B}$. 
{Such a scale-{invariant} behavior of $T^*\propto \mu$} {implies that three-body correlations are still alive in the high-density regime.}
{For comparison, {in Fig.~\ref{fig1},} we plot the {points} where {the QMC result~\cite{McKenny2} for} the {isothermal} compressibility $\kappa$ 
{normalized by the ideal-gas value $\kappa_0$ is minimal with respect to $\mu$.}
Interestingly, {these points coincide well} with the {$T^*$-$\mu$ relation} at low temperature.
{In QCD,} the sound velocity, which is proportional to $\kappa^{-1/2}$ at $T=0$, {is predicted to be peaked in} 
the hadron-quark crossover {regime}~\cite{BaymR,KojoR}.
{Thus,} our results suggest that such macroscopic behavior manifests the emergence of Cooper triples in {both systems (for details, see Supplemental Material S4~\cite{SupMat}).}}
\par
\noindent {\it Conclusion---}
We have clarified {the conditions of temperature and chemical potential that {allow}} Cooper triples and {trimers {to} occur in the {1D} equilibrated system} of three-component fermions with three-body attraction. 
{We have found a} non-trivial crossover from {the} tightly bound trimer {phase} to the Cooper triple phase with increasing chemical potential, {which is analogous to the hadron-quark crossover in QCD.} 
{The characteristic temperature {$T^*$ of} Cooper triples agrees {well} with the compressibility {minima of the QMC result in {this 1D} system,} implying that 
{{the hadron-quark crossover} is accompanied by the emergence of quark Cooper triples.}
Indeed, this scenario is {physically analogous to}
 McLerran-Reddy model {for quarkyonic matter}~\cite{McLerranReddy}.}

{For future perspectives, the comparison of the compressibility between our diagrammatic approach and {the existing QMC result}
would be helpful to confirm {the relevance of Cooper triples.}     
Since the deconfined phase near the $T^*$ minimum is dominated by strong fluctuations~\cite{Tajima1}, the compressibility anomaly could not be understood by usual {quasiparticle} pictures. }
It is also interesting to address quartet {condensation}~\cite{Kamei,Sogo}, {dual bosonic systems~\cite{Valiente3,Valiente4},
higher dimensions~\cite{Akagami}}, and lattice systems~\cite{DelRe,Christianen}.
{Moreover, the three-body loss can be a useful probe for the emergence of Cooper triples as in the case of Efimov effects~\cite{Naidon,Tajima2021}.}
\par
{We are grateful to} {Y. Hidaka for reading the manuscript and giving us pertinent comments, and}
 M. Horikoshi, T. Hatsuda, H. Yabu, E. Nakano, J. Takahashi, K. Nishimura, T. Hata, K. Ochi, and S. Akagami for useful discussion.
This work is supported by Grants-in-Aid for Scientific Research {provided by} JSPS {through Nos.\ 18H01211, 18H05406, and 20K14480.}
S.T. was supported by the RIKEN Special Postdoctoral Researchers Program. 



\clearpage
\section{Supplementary Material}
\beginsupplement

\setcounter{equation}{0}
\setcounter{figure}{0}
\setcounter{table}{0}
\setcounter{page}{1}
\makeatletter
\renewcommand{\theequation}{S\arabic{equation}}
\renewcommand{\thefigure}{S\arabic{figure}}
\renewcommand{\bibnumfmt}[1]{[S#1]}
\renewcommand{\citenumfont}[1]{S#1}

\section{S1. Atom-trimer resonance model for a tunable three-body interaction}
\begin{figure}[h]
\begin{center}
\includegraphics[width=6cm]{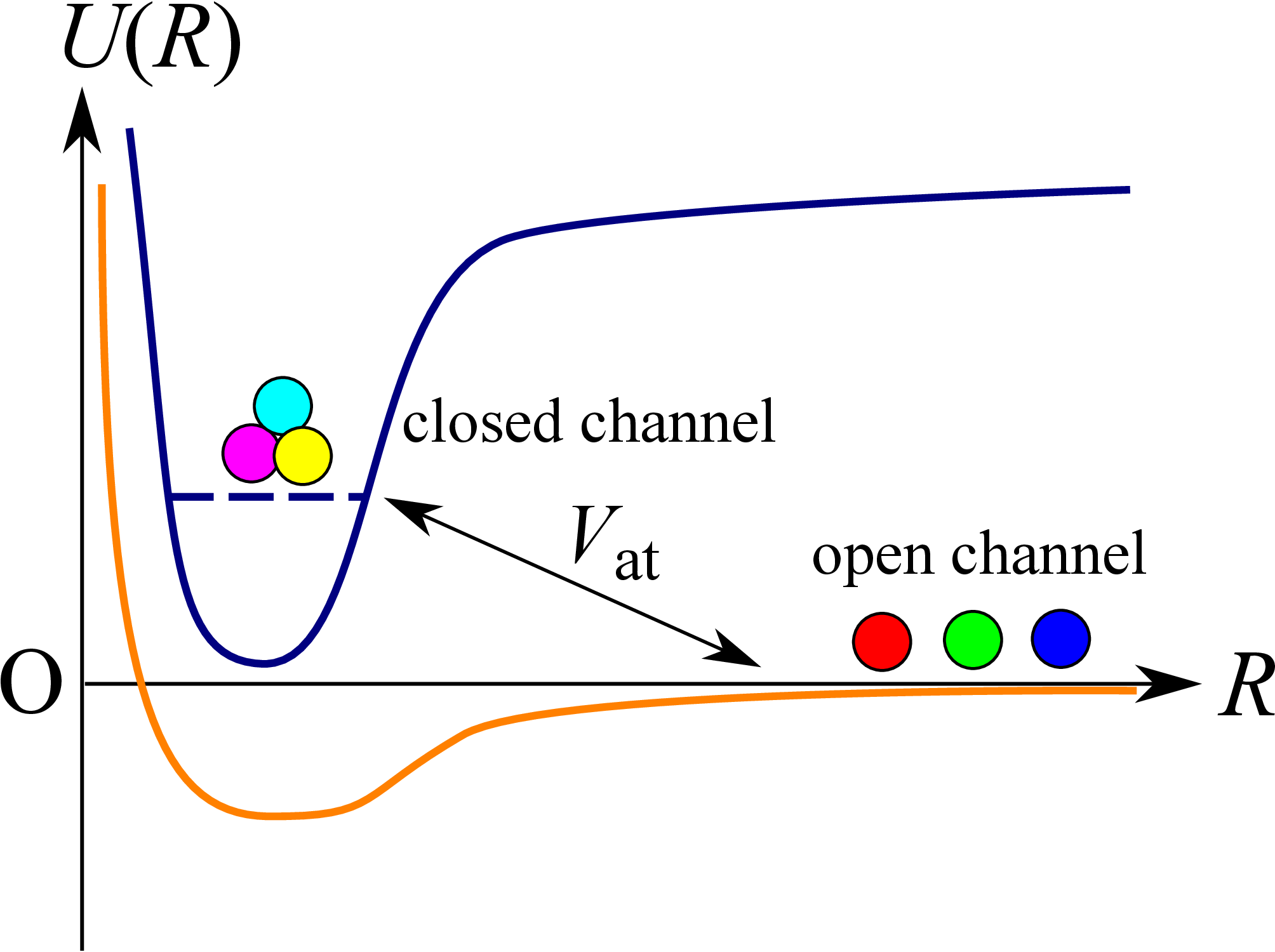}
\end{center}
\caption{Schematic {illustration of} an atom-trimer resonance model.
$U(R)$ and $R$ are the three-body potential and the hyperradius, respectively, {while}
$V_{\rm at}$ {denotes} the coupling between the open channel atoms and the closed channel trimer.
}
\label{figS1}
\end{figure}
{In this supplement, we illustrate how}
the three-body interaction among three-state fermions  can be realized.
{Our} basic idea, {which requires coupling between an open channel continuum state 
and an excited bound trimer state, is summarized} in Fig.~\ref{figS1}.
{For} such an atom-trimer resonance model, we {develop a} coupled-channel formalism. 
The corresponding Hamiltonian reads
\begin{align}
H_{\rm at}&=\sum_{\bm{p},\gamma}\varepsilon_{\bm{p},\gamma}c_{\bm{p},\gamma}^\dag c_{\bm{p},\gamma}
+\sum_{\bm{P}}\left(\varepsilon_{\bm{P}}^{\rm t}+\nu\right)A_{\bm{P}}^\dag A_{\bm{P}}
\cr
+&V_{\rm at}\sum_{\bm{P},\bm{k},\bm{q}}\left( A_{\bm{P}}^\dag c_{\frac{\bm{P}}{3}-\bm{k}-\frac{\bm{q}}{2},{\rm b}}c_{\frac{\bm{P}}{3}+\bm{q},{\rm g}} c_{\frac{\bm{P}}{3}+\bm{k}-\frac{\bm{q}}{2},{\rm r}} +{\rm h.c.}\right), \nonumber \\
\end{align}
where $\varepsilon_{\bm{P}}^{\rm t}=P^2/(6m)$ and $A_{\bm{P}}^{(\dag)}$ are the kinetic energy and the annihilation (creation) operator of the closed channel trimer with the energy level  $\nu$, respectively.
Here, we have considered the equal mass {for all types of fermions} and defined $\varepsilon_{\bm{p},\gamma}=|\bm{p}|^2/(2m)$.
For simplicity, we ignore other background interactions {by assuming} that the system is far away from ordinary magnetic Feshbach resonances.
\par
The atom-trimer coupling $V_{\rm at}$ may occur through the hyperfine interaction or the optical transition as in the case of the optical Feshbach resonance.
The closed-channel trimer state would be found in few-body or quantum chemical calculations~\cite{Cvitas,Ghassemi,Yan}, as well as {in} future precise spectroscopic experiments~\cite{Horikoshi}.
In this model, the three-body $T$-matrix $T_{3}(\bm{P},\Omega_+)$ is given by
\begin{align}
T_{3}(\bm{P},\Omega)&=\frac{V_{\rm at}^2}{\Omega_+-\varepsilon_{\bm{P}}^{\rm t}-\nu-V_{\rm at}^2\Xi_0(\bm{P},\Omega)},
\end{align}
where $\Xi_0(\bm{P},\Omega_+)$ {is the bare three-body propagator. }
\par
{Before considering the one-dimensional case of interest here, we first consider the
bare three-body propagator in three dimensions, which} can be obtained as
\begin{align}
\Xi_0(\bm{P},\Omega_+)&=\sum_{\bm{k},\bm{q}}\frac{1}{\Omega_+-\varepsilon_{\frac{P}{3}-\bm{k}-\frac{\bm{q}}{2},{\rm r}}-\varepsilon_{\frac{\bm{P}}{3}+\bm{q},{\rm g}}-\varepsilon_{\frac{\bm{P}}{3}+\bm{k}-\frac{\bm{q}}{2}}}\cr
&=-\frac{m}{12\sqrt{3}\pi^3}\left[\Lambda^2\left(m\Omega-\frac{\bm{P}^2}{6}+\frac{\Lambda^2}{2}\right)\right.\cr
&+\left.\left(m\Omega-\frac{\bm{P}^2}{6}\right)\ln\left(\frac{\Lambda^2+\bm{P}^2/6-m\Omega_+}{\bm{P}^2/6-m\Omega_+}\right)\right].
\end{align}
{At $\bm{P}=\Omega_+=0$, therefore,} we obtain the three-body $T$-matrix {in three dimensions} as 
\begin{align}
\label{eq:t3-3d}
T_{3, {\rm 3D}}(\bm{0},0)&=\left(-\frac{\nu}{V_{\rm at}^2}+\frac{m\Lambda^4}{24\sqrt{3}\pi^3}\right)^{-1}\cr
&\equiv-\frac{V_{\rm at}^2}{\nu_{\rm R}},
\end{align}
where the renormalized trimer energy level $\nu_{\rm R}$ is defined as
\begin{align}
\nu_{\rm R}=\nu-\frac{m\Lambda^4}{24\sqrt{3}\pi^3}V_{\rm at}^2.
\end{align}
{Equation} (\ref{eq:t3-3d}) indicates that {a} bound {trimer} state {can} appear {at $\nu_{\rm R}\simeq0$}, {while} the three-body coupling can be changed by tuning $\nu_{\rm R}$ as in the case of the magnetic Feshbach resonance {in which} the renormalized closed channel molecular energy is tuned~\cite{ChinS}.
\par
{In one dimension,} the present coupled channel model can be reduced to the single-channel model
when $V_{\rm at}$ {and $\nu$ become} sufficiently large {in such a way as to keep $V_{\rm at}^2/\nu$ finite. 
In fact}, $T_3(P,\Omega_+)$ is given by
\begin{align}
\label{eq:t3-1d_2c}
T_3(P,\Omega_+)&=\left[\frac{\Omega_+-P^2/(6m)-\nu}{V_{\rm at}^2}+\frac{m}{2\sqrt{3}\pi}\right.\cr
&\times\left.\ln\left(\frac{\Lambda^2+P^2/6-m\Omega_+}{P^2/6-m\Omega_+}\right)
\right]^{-1}, 
\end{align}
{which reduces to} Eq.~(\ref{eq:t3-1d}) {when one sets $g_3=-V_{\rm at}^2/\nu$ and 
$|\nu| \gg |\Omega_+-P^2/(6m)|$.}

\par
For more realistic situations, we have to consider a light-induced loss {in the presence of} an optical transition between open and closed channels.
Recently, {however,} the ``dark-state" optical method has been proposed to avoid such a {loss}~\cite{Wu1,Wu2}.
Indeed, the optical control of a scattering length and an effective range in two-body scattering is experimentally {feasible for} a $^6$Li Fermi gas~\cite{Jagannathan,Arunkmar1,Arunkmar2}.
{Application} of this method to the present atom-trimer resonance is left for future work.  


\section{S2. Properties of three-body bound state in vacuum}
{In this supplement, }{we summarize the properties of the three-body bound state in vacuum.}
One can obtain the renormalization group flow of $g_3$ as
\begin{align}  
\label{eq:RG}
\frac{\partial g_3}{\partial\ln\lambda}=\frac{m}{\sqrt{3}\pi}g_3^2,
\end{align}
indicating the asymptotic freedom, {together} 
with the emergence of {an} additional {momentum scale $\Lambda$ after the integration 
with respect to the momentum scale $\lambda$ at which one probes $g_3$.  Then, after}
analytically performing the momentum integration in $T_3(P,\Omega_+)$ {in vacuum} and
finding the negative energy pole, we obtain
\begin{align}
\label{eq:t3-1d}
T_3(P,\Omega_+)&=\left[\frac{1}{g_3}+\frac{m}{2\sqrt{3}\pi}\ln\left(\frac{\Lambda^2+P^2/6-m\Omega_+}{P^2/6-m\Omega_+}\right)\right]^{-1}\cr
&\simeq\frac{2\sqrt{3}\pi}{m}\left[\ln\left(\frac{mE_{\rm B}}{P^2/6-m\Omega_+}\right)\right]^{-1},
\end{align}
where
\begin{align}
\label{eq:eb}
E_{\rm B}=\frac{\Lambda^2}{m}e^{\frac{2\sqrt{3}\pi}{mg_3}} 
\end{align}
is the three-body binding energy {in vacuum}, which is consistent with {that derived in} Ref.~\cite{sDrut1}.
In the second line of Eq.~(\ref{eq:t3-1d}), we assumed $\Lambda \gg \sqrt{mE_{\rm B}}$. 
{We remark that one} can derive Eq.~(\ref{eq:RG}) from the condition $\frac{\partial E_{\rm B}}{\partial\ln \Lambda}=0$, i.e., $E_{\rm B}$ in Eq.~(\ref{eq:eb}) does not explicitly depend on $\Lambda$ after the renormalization of $g_3$.

\section{S3. Saha-Langmuir equation for a fermion-trimer mixture}
{In this supplement, we derive the {degree of dissociation} $\alpha\equiv\rho_{\rm f}/\rho$ and the 
dissociation temperature $T_\alpha$ in} a classical fermion-trimer mixture {of total fermion number density $\rho$ and free fermion number density $\rho_{\rm f}$
from} the Saha-Langmuir equation, 
\begin{align}
\frac{\rho_{\rm f}^3}{\rho-\rho_{\rm f}}=3^{3/2}\frac{mT}{2\pi}e^{-E_{\rm B}/T},
\end{align}
{which can in turn be obtained from}
\begin{align}
\rho_{\rm f}&=3\int\frac{dp}{2\pi}\exp\left[-\frac{p^2/(2m)-\mu}{T}\right]\cr
&=3z\sqrt{\frac{m T}{2\pi}}
\end{align}
and
\begin{align}
\rho-\rho_{\rm f}&=3\int\frac{dP}{2\pi}\exp\left[-\frac{P^2/(6m)-E_{\rm B}-3\mu}{T}\right]\cr
&=3z^3e^{E_{\rm B}/T}\sqrt{\frac{3mT}{2\pi}},
\end{align}
{where} $z=e^{\mu/T}$ is the fugacity.
{In terms of the fugacity, which is supposed to be sufficiently small, we can write} the {degree of dissociation} as
\begin{align}
\label{eq:alpha}
\alpha
=\frac{1}{1+\sqrt{3}z^2e^{E_{\rm B}/T}}.
\end{align}
{By solving Eq.~(\ref{eq:alpha}) with respect to $T$ and setting the resultant $T$ as $T_{\alpha}$, 
one finally obtains}
\begin{align}
T_\alpha=\frac{2\mu+E_{\rm B}}{\ln\left(\frac{1-\alpha}{\sqrt{3}\alpha}\right)}.
\end{align}

\begin{figure}[t]
\begin{center}
\includegraphics[width=6cm]{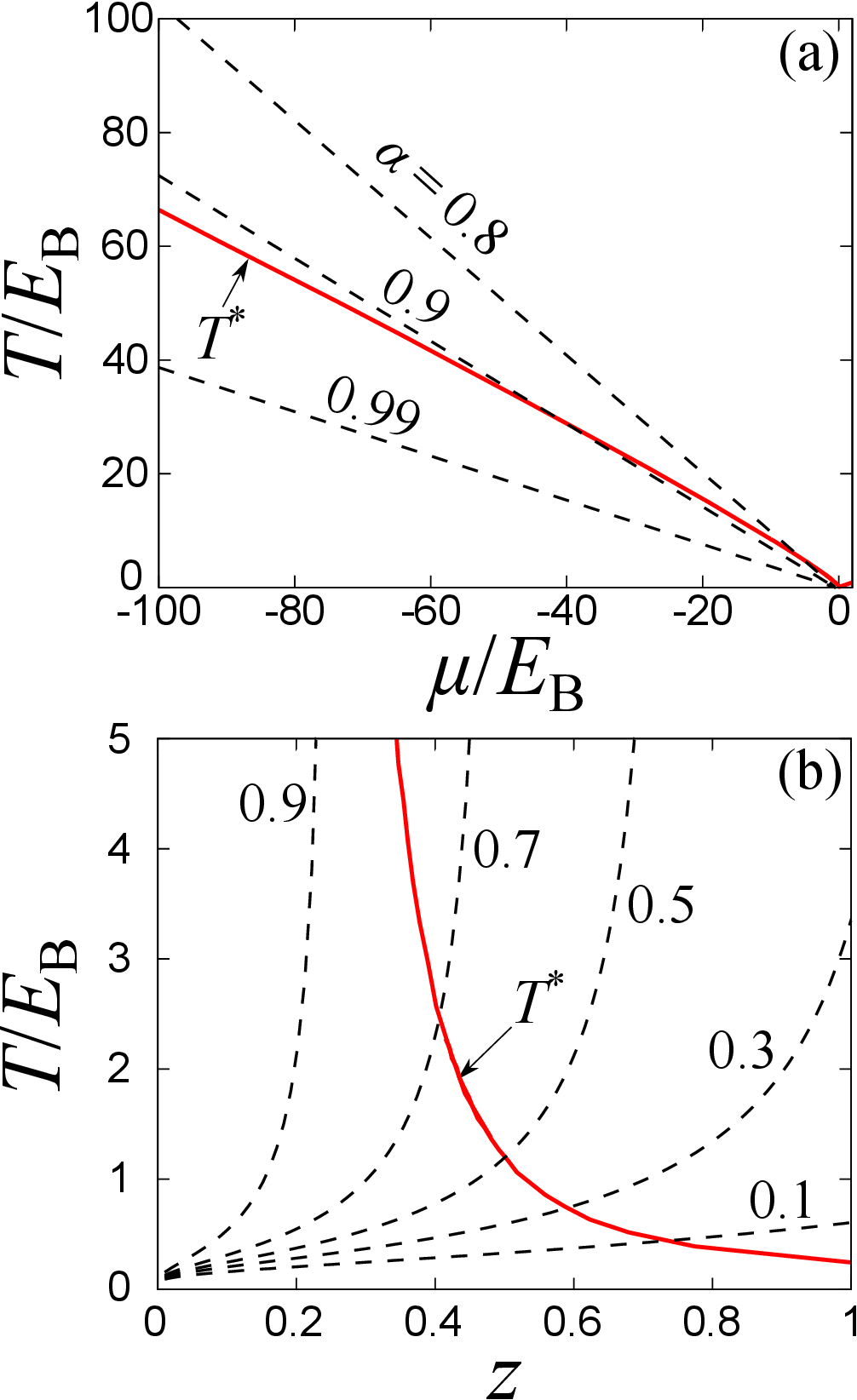}
\end{center}
\caption{{The dissociation temperatures $T_{\alpha}$ as functions of (a) the chemical potential $\mu/E_{\rm B}$ and (b) the fugacity $z=e^{\mu/T}$ in the low-density regime.
The {number} alongside each dashed line {denotes} the degree of dissociation $\alpha$.}
}
\label{figS2}
\end{figure}
{
{To clarify the role played by the dissociation temperature $T_{\alpha}$, Eq.\ (6), in explaining the low-density behavior of $T^{*}$, we have drawn} Figs.~\ref{figS2}(a) and (b) {in which $T_{\alpha}$ is plotted as a function of the chemical potential $\mu$ and the fugacity $z$, respectively,} in the low-density regime.  One can see {from Fig.\ \ref{figS2}(a)} that $T_{\alpha=0.9}$ is close to $T^*$ in the sufficiently low-density regime ($\mu/E_{\rm B}\lesssim -10 $).  {This suggests that thermal dissociation in terms of $\alpha$ gives a valid picture of the trimer phase at finite temperature up to $\sim T^{*}$.}}
\par
{
{Then, let us proceed to consider the behavior of $T_\alpha$ in the $T$-$\mu$ plane as shown in Fig.\ 1.  According to this behavior, $\alpha$ decreases with increasing temperature at given $\mu$.  To understand such a seemingly counterintuitive result,} it is useful to see the fugacity dependence of $T_{\alpha}$ shown in Fig.~\ref{figS2}(b).
{One can find from this figure that at given $z$,} $\alpha$ monotonically {increases} with increasing temperature.  
This tendency {can be easily understood from Eq.~(\ref{eq:alpha}). }
{We thus conclude that the decrease in $\alpha$ with increasing temperature} at given $\mu$ is due to the {simultaneous} enhancement of $z$
{and that the comparison between $T^*$ and $T_\alpha$ as shown in Fig.~1 is still meaningful.}} 
\par
{{Strictly speaking,} the Saha-Langmuir equation becomes {no longer} valid in the region where {the thermal} medium {leads the trimer binding to vanish} ($T\gesim T^*$) since the robust trimer binding is {taken for granted} in this equation.  The {estimate} of $\alpha$ {from the Saha-Langmuir equation} is {nevertheless} useful to {roughly know} how dissociated fermions and trimers are distributed in the deep inside of the tightly bound trimer phase ($T\lesssim T^*$). 
{According to such an estimate at largely negative chemical potential,} the region near the vacuum {where the fugacity is vanishingly small} is dominated by dissociated fermions, {while} the fraction of the tightly bound {trimers, $1-\alpha$,} increases {up to about 0.1} around $T^*$, as shown in Fig.~\ref{figS2}(a). {Note, however, that}
the medium effect {has to} suppress such a trimer fraction above $T^*$.
}

\section{S4. Ground-state properties based on {a} McLerran-Reddy-like model for the Cooper {triple} state}
{In order to see the role of Cooper triples on the behavior of the isothermal compressiblity,
we consider a simple 1D effective model of the fermion-triple mixture where the total density is given by}
\begin{align}
\label{eq:S4-1}
\rho=3\rho_{\rm f,0}+3\rho_{\rm C}.
\end{align}
In Eq.~(\ref{eq:S4-1}), we have defined the number density of an ideal Fermi gas {as}
\begin{align}
{\rho_{\rm f,0}=\frac{(2m\mu)^{\frac{1}{2}}}{\pi}}
\end{align}
and the {number density} of degenerate composite fermions (i.e., Cooper triples) {as}
\begin{align}
{\rho_{\rm C}=\frac{(2m)^{\frac{1}{2}}}{\pi}(3\mu+E_{\rm B}^{\rm M})^{\frac{1}{2}},}
\end{align}
In a non-relativistic system at $T=0$, the sound velocity $c_{\rm s}$ can be obtained from
\begin{align}
c_{\rm s}^2= \frac{1}{m\rho \kappa},
\end{align}
where
\begin{align}
\kappa = \frac{1}{\rho^2}\left(\frac{\partial \rho}{\partial \mu}\right)
\end{align}
is the compressibility.
{Obviously,} the sound-velocity maximum is deeply related to the minimum of $\kappa$.
One can analytically obtain
\begin{align}
\label{eq:k1d}
{\kappa= \frac{3(2m)^{\frac{1}{2}}}{2\pi\rho^2}\left[\frac{1}{\sqrt{\mu}}+\frac{1}{\sqrt{3\mu+E_{\rm B}^{\rm M}}}\left(3+\frac{\partial E_{\rm B}^{\rm M}}{\partial \mu}\right)\right].}
\end{align}
{In particular, the dimensionless quantity $\kappa/\kappa_0$ calculated in Ref.~\cite{McKenny2} ($\kappa_0$ is the ideal-gas value) is given by}
\begin{align}
{\frac{\kappa}{\kappa_0} =1+\sqrt{\frac{\mu}{3\mu+E_{\rm B}^{\rm M}}}\left(3+\frac{\partial E_{\rm B}^{\rm M}}{\partial \mu}\right), }
\end{align} 
which indicates that the in-medium three-body binding energy $E_{\rm B}^{\rm M}$ plays a crucial role in the behavior of {$\kappa/\kappa_0$ and thus} $c_{\rm s}$.
{First of all, it is important to note that} $E_{\rm B}^{\rm M}$ is a decreasing function of $\mu$ at low temperature.
{Then, the term $\frac{\partial E_{\rm B}^{\rm M}}{\partial \mu}(\leq 0)$ acts to decrease $\kappa$.  Secondly, 
it is to be noted that} $E_{\rm B}^{\rm M}$ {sharply} decreases around {the crossover region, i.e.,} $\mu\simeq 0.05E_{\rm B}$.
{This behavior leads to} the minimum of $\kappa$ {in the crossover regime}. 
{Although this minimum could} be negative {and hence indicate a} breakdown of {the present} simple model,
one can {qualitatively} understand from this analysis how {crucial} the Cooper triple formation is for the sound-velocity {peak} in the crossover regime, 
{which is reminiscent of} quarkyonic matter.
\par
{We remark that the isothemal compressibility has the same tendency in 3D.
In fact, it is given by
\begin{align}
\kappa_{\rm  3D}= \frac{3(2m)^{\frac{3}{2}}}{4\pi^2\rho^2}\left[\sqrt{\mu}+\sqrt{3\mu+E_{\rm B}^{\rm M}}\left(3+\frac{\partial E_{\rm B}^{\rm M}}{\partial \mu}\right)\right],
\end{align}
where the term $\frac{\partial E_{\rm B}^{\rm M}}{\partial \mu}$ appears in the same way as Eq.~(\ref{eq:k1d}).
Thus, the emergence of the Cooper triples is a possible origin of the sound velocity in
the hadron-quark crossover.
Indeed, this simple model based on the fermion-trimer mixture is similar to the McLerran-Reddy model for quarkyonic matter in Ref.~\cite{MR}.
}



\begin{thebibliography}{98}
\bibitem{note1} 
When we perform the momentum integration in Eq.~(\ref{eq:xi}), we use a transformation $\bm{q}\rightarrow\frac{2}{\sqrt{3}}\bm{q}$.
Although it modifies the {momentum cutoff $\Lambda$ that appears in the expression for} $E_{\rm B}$,  
{by a factor of order unity, $\ln(mE_{\rm B}/\Lambda^2)$} is not changed {up to $O(g_3^{-1})$}.
\end{thebibliography}
\end{document}